\documentclass[11pt]{article}
\usepackage{pdproc,epsfig,graphicx} 
\usepackage{nicefrac}

\begin{document}

\renewcommand{\thefootnote}{\alph{footnote}}
  
\title{HOW MUCH CAN WE LEARN FROM SN1987A EVENTS?\\
{\small Or: An Analysis with a Two-Component Model for the Antineutrino Signal}}

\author{FRANCESCO VISSANI}

\address{INFN, Laboratori Nazionali del Gran Sasso, Assergi (AQ), Italy\\
 {\rm E-mail: vissani@lngs.infn.it}}

  \centerline{\footnotesize and}

\author{GIULIA PAGLIAROLI}

\address{INFN, Laboratori Nazionali del Gran Sasso, Assergi (AQ), Italy\\
L'Aquila University, Coppito (AQ), Italy\\
 {\rm E-mail: giulia.pagliaroli@lngs.infn.it}}

\abstract{We analyze the data of Kamiokande-II, IMB, Baksan 
using a parameterized description of the antineutrino
emission, that includes an initial phase of intense luminosity. 
The luminosity curve, the average energy of $\bar\nu_e$
and the astrophysical parameters of the model, derived by fitting 
the observed events (energies, times and angles) are 
in reasonable agreement with the generic expectations
of the delayed scenario for the explosion.}
   
\normalsize\baselineskip=15pt

\section{Introduction}

We begin with a rapid historical excursus, 
with emphasis on the issues that are 
relevant for data analysis.

\begin{itemize}

\item
Colgate \& White 1966\cite{cg} 
propose the paradigm for the explanation 
of the core collapse supernovae,
where neutrinos are the key agents.

\item
Nadyozhin 1978\cite{nad} 
concludes a detailed calculation that demonstrates an initial
phase of intense neutrino luminosity.

\item
Bethe \& Wilson 1985\cite{bw} 
suggest that the energy deposition 
on a scale of half a second can re-energize the stalled shock wave.

\item
1987: Kamiokande-II\cite{kii}, IMB\cite{imb}, Baksan\cite{bak} 
and LSD\cite{lsd} observe 
several events in correlation with SN1987A.

\item
Several authors--e.g., Bahcall 1989\cite{bah}--remark 
that non-LSD data generically meet the expectations. 
Then the main interest shifts on the relevance of oscillations.

\item
Lamb \& Loredo 2002\cite{ll} (LL) discuss whether SN1987A data indicate 
specific imprints of the delayed scenario\footnote{With the term 
`delayed scenario' we refer here and in the following to the scenario
for the explosion put forward by Bethe and Wilson, that incorporates
the initial phase of intense neutrino luminosity discussed by
Nadyozhin (this is also called 
`standard scenario' or `neutrino assisted explosion').}
such as the initial phase of intense luminosity.

\item
Imshennik~\&~Ryazhskaya~2004\cite{ir} suggest a 2 stage scenario 
with essential role of rotation and possible explanation of LSD
events detected 4.5 hours earlier.

\end{itemize}
Next we summarize the present status: 
although there is mounting evidence that the delayed scenario is 
correct, a conclusive proof is still missing; while most people
is convinced that SN1987A neutrinos confirmed the general picture of 
the explosion, there are annoying doubts on whether 
SN1987A was a standard object or not; while we continue sharpening our 
theoretical tools, we hope intensely in a Galactic supernova event 
to progress in our understanding.

In short, one should be aware that until 
the theoretical picture will be definitively assessed,  
the interpretation of the results from SN1987A 
will continue to contain elements of uncertainty.
But if we want to test the expectations, 
we are forced to answer the question:
{\em What are the generic expectations for the delayed scenario?}
We summarize the features that are relevant for our analysis:
\begin{itemize}
\item 
In conventional detectors
(scintillators and water Cherenkov) 
the main detection reaction 
is the inverse-beta decay of electron antineutrinos 
on free protons (IBD):
$\bar\nu_e p\to n e^+ $.

\begin{figure}[t]
$$\includegraphics[width=0.5\textwidth]{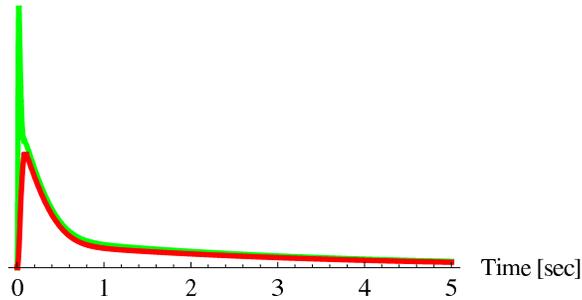}$$ 
\caption{\it 
Sketch of $\nu_e$ and $\bar\nu_e$ (higher and lower)
luminosity curves. The excess of $\nu_e$ in the first 50 ms 
(`neutronization' phase) gives 
a small contribution to the total number of 
observable events.
\label{fig1}}
\end{figure}

\item
The main part of the emission happens in two stages
as shown in Fig.~\ref{fig1}.
10-20\%  of the energy is radiated in an early 
phase, here called {\bf accretion}, that should last 
about half-a-second; 80-90\% of the energy is radiated later, during
the phase of neutron star {\bf cooling}, namely in a quiet thermal phase.

\item
The main reactions for energy radiation during accretion are
$e^- p\to n \nu_e$ and $e^+   n\to p \bar\nu_e$, the second being 
the inverse of IBD. The presence around the nascent neutron star 
of an abundant amount of $e^+e^-$ plasma 
ensures a large flux of  $\nu_e$ and $\bar\nu_e$, that are key 
ingredients to revive the stalled shock wave.
[See Janka\cite{j} for a review.]

\item
During the cooling phase, neutrinos of all 
species 
($\nu_e,\nu_\mu,\nu_\tau,\bar\nu_e,\bar\nu_\mu,\bar\nu_\tau$)
are radiated with 
similar luminosities; this feature 
is occasionally called `equipartition'.

\end{itemize}
In this talk, based on a work in collaboration with
M.L.~Costantini and A.~Ianni\cite{paglia},
we present an analysis of SN1987A data along these lines.
We will compare two models for neutrino emission, namely 
the conventional one, when the neutrino emission is 
described by a `one-component' (cooling)
model, and a `two-component' (cooling+accretion) model which 
resembles more closely the expectations of the delayed scenario.
Preliminary results have been documented in\cite{mosca,prep,ifae,brasile};
see also\cite{jcap}, in particular Sect.~3.2 there.

\section{Analysis of SN1987A Observation with One-Component Model}

\subsection{The Conventional Model for $\bar\nu_e$ Emission 
(=Exponential Cooling)}
This is the model adopted in most of the quantitative analyses 
of SN1987A events (the notable exception, to be discussed later, 
being the work of LL\cite{ll}).
The thermal emission of the neutron star
is parameterized by a black body model with a 
steadily decreasing temperature:
\begin{equation}
\left\{
\begin{array}{l}
\displaystyle \frac{dN_c}{dt\; dE}=\frac{R_c^2}{2\pi} \cdot
\frac{E^2}{1+\exp[E/T_c(t)]}\\[2ex]
\mbox{with }T_c(t)=T_c\cdot \exp[ -t/(4\tau_c)]
\end{array}
\right.
\end{equation}
where the suffix $c$ means `cooling' and where we use 
for convenience the natural units $\hbar=c=k_B=1$.
The 3 parameters of the model are:
\begin{enumerate}
\item 
the neutrinosphere radius $R_c$ that describes the 
{intensity} of the emission;
\item 
the initial temperature $T_c$
that fixes the {average energy} of $\bar\nu_e$;
\item 
the luminosity time scale $\tau_c$
that quantifies the duration of the process.
\end{enumerate}
An isotropic emission from $D=50$ kpc is assumed so that the 
flux is simply $\Phi_c=1/(4\pi D^2)\times dN_c/dEdt $. For obvious 
reasons, this model for the antineutrino emission
is called `exponential cooling'.

\subsection{Procedure of Analysis}
Next we describe the procedure of analysis we adopted. 
It is significantly more complex than the usually 
followed procedures 
(though we verified a posteriori none of these 
technical improvements leads to qualitative changes in the 
conclusion) and elaborates on the results obtained by LL\cite{ll}:\\
a) Since the absolute times of Kamiokande-II and Baksan are not
measured precisely, and the time between the first neutrino and the 
first event in IMB is unknown, we use only the relative 
times of the events ($t_1=0$ for any detector).\\
b) Since the duration of the signal is not known a priori, 
we consider the data in a unified time window
of 30 s.\\
c) We account for the measured background (Kamiokande-II and Baksan), 
dead-times and angular-bias (IMB).\\
d) We account for the error in the energy measurement 
in all detectors.
More details of the analysis are described in Appendices A and B.\\
The expected number of signal events obtained from the differential rate
\begin{equation}
\frac{dN}{dt dE_e d\cos\theta}= N_p \frac{d\sigma_{\bar\nu_e
    p}}{d\cos\theta}(E_{\nu},\cos\theta ) 
\Phi_{\bar\nu_e}(t, E_{\nu}) 
 \xi_d(\cos\theta) \eta_d(E_e) \frac{dE_{\nu}}{dE_e},
\end{equation}
that requires to know the number of protons in the detector, 
the interaction (IBD) cross section, the $\bar\nu_e$ flux, 
the angular bias, the detection efficiency and the Jacobian.

\subsection{Results}

A straightforward fit yields the following values of the 
three astrophysical parameters
\begin{equation}
R_c=26\mbox{ km},\ T_c=4.5\mbox{ MeV},\ \tau_c=3.9\mbox{ s}\
\end{equation}
while the best fit of the offset times are zero, due to the
fact that the signal is bounded to decrease with time.
Our best fit is very close to the value in Bahcall book.\cite{bah}

The difference with LL is mostly due to their 
treatment of efficiency--not of  
background, inclusion of angles,  
or better cross section.
Indeed, LL include efficiency in the exponential of 
$e^{\!-\!\sum \mu_j(\rm with)} \prod \mu_i(\rm w/o)$, 
we include it in both terms.
A formal justification of our procedure 
is given in Appendix~A.

\subsection{Should We Stop Here?}

\begin{figure}[t]
$$\includegraphics[width=0.4\textwidth,
height=0.4\textwidth,angle=270]{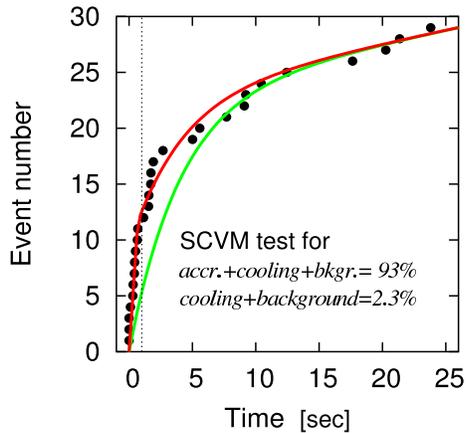}$$ 
\caption{\footnotesize\it Temporal distribution of the events; the end 
of the first second is marked by a dashed line.
\label{h}}
\end{figure}

At this point we face the question: should we be content of the
generic agreement with the expectation and stop the analysis here?
One could presume that Bahcall would have answered ``yes''.
Indeed, in his book,\cite{bah} the analysis of SN1987A data is 
introduced by the statement:
\begin{quote}
{\em Unfortunately, a ``minimum'' model has proven adequate 
to describe the sparse amount of data.}
\end{quote}
The study of the `exponential cooling' model 
is then commented with the sentence:
\begin{quote}
{\em The success of this simplified ``standard''
model suggests that it will be difficult to use the neutrino events
observed from SN1987A to establish more detailed models. }
\end{quote}
We wish to note that ``difficult'' does not mean
  ``impossible'' and, 
remaining aware of this authoritative opinion, we would like to 
continue the discussion begun by Lamb \& Loredo: 
{\em whether it is worthwhile to go beyond  the exponential cooling model.}

Actually, observations offer us some motivation to proceed. In fact, 
there is a hint of an increased luminosity in the first second.
This is evident from Fig.~\ref{h}, where we show the temporal
distribution of the observed events, from\cite{mosca} (see also\cite{jcap}). 
For comparison, we show also  
two time distributions, comprising also 7 background events
(the expectations is 5.6 in Kamiokande-II and 1 in Baksan).
In the lower one, all 22 signal events belong to the 
cooling component.
In the upper one, only 13 signal events belong to cooling;
the remaining 9 belong to the accretion component.
In the figure it is indicated the GOF of these two hypotheses,
evaluated with a Smirnov-Cram\`er-Von Mises test.

\section{Analysis of SN1987A Observation with Two-Component Model}

\begin{figure}[t]
\hskip1cm
\includegraphics[width=0.35\textwidth,
height=0.35\textwidth]{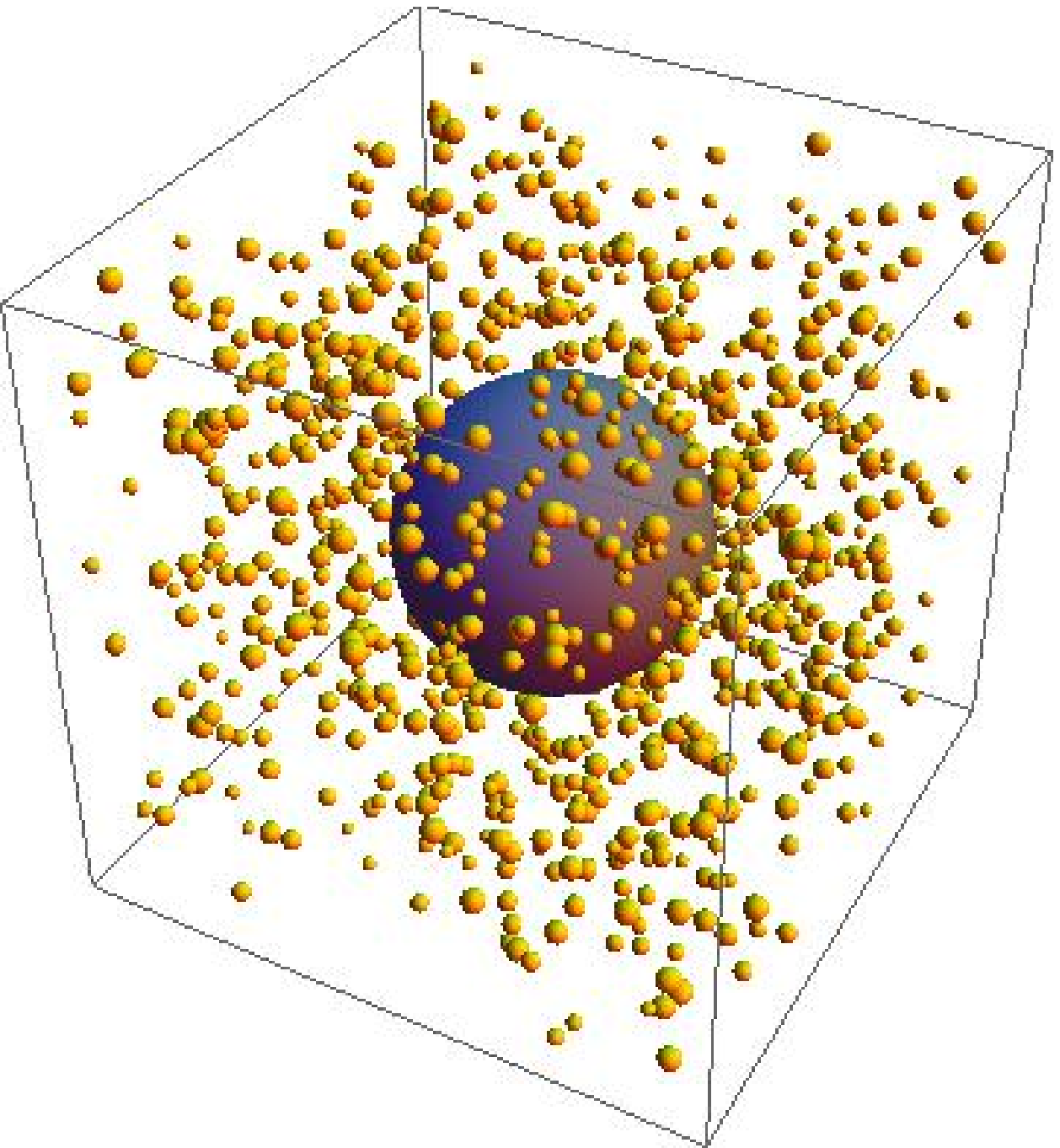}~\vskip-4.5cm
$$
\ \ \ \ \ \ \ \ \ \ \ \ \ \ \ \ \ \ \ \ \ \ \  \ \ \ \ \ \ \ \ 
\  \ \ \ \ \ \ \ \ 
\begin{array}{l}
\mbox{\tt Thermal $e^+$'s react} \\[-.8ex]
\mbox{\tt on target neutrons} \\[0ex]
\ \ \ \displaystyle N_n(t)=\frac{Y_n\  M_a(t)}{m_n}\\[-0.2ex]
\mbox{\tt yielding many $\bar\nu_e$'s.} 
\end{array}
$$
\vskip1.5cm
\caption{\footnotesize\it The yellow (gray) centers surrounding the NS
(dark gray) represent pictorially the individual 
reactions $e^-p\to n \nu_e $ and $e^+n\to p \bar\nu_e $ 
occurring during accretion.\label{picco}}
\end{figure}

\subsection{How to Describe the Initial Emission (=Accretion)}

The initial emission is  dominated by 
the quasi-transparent accreting region: 
on top of the previous black body emission, 
we simply need to model the radiation  of $\bar\nu_e$ from 
$e^+n\to p \bar\nu_e$. 
Quite directly, we use:
\begin{equation}
\frac{dN_a}{dtdE}= 
\frac{1}{\pi^2} N_n(t) \sigma_{e^+n}(E_\nu)  \frac{E_e^2}{1+\exp[E_e/T_a(t)]}
\end{equation}
The conceptual scheme is represented pictorially in Fig.~\ref{picco},
where the second astrophysical parameter $M_a$ (the accreting mass) 
besides the temperature of the positron plasma, is introduced. 
The assumed description of how $T_a$ and $M_a$ evolve in time 
is given in Appendix B, but it should be not a surprise that this
description brings in a third parameter, $\tau_a$, the duration of the
process of accretion. We compare in the next page 
our description of $\bar\nu_e$ emission 
with the one suggested by Lamb and Loredo.

\begin{figure}[hb]
$$\includegraphics[width=0.30\textwidth,height=0.45\textwidth,angle=270]{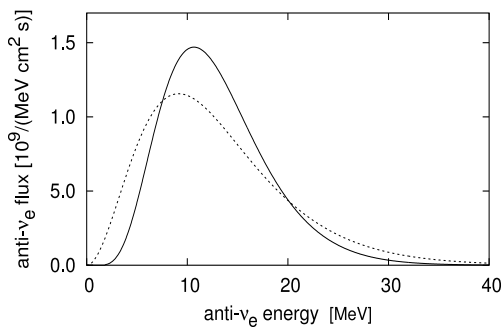}$$
\caption{ \footnotesize \it
Continuous curve: 
$\bar\nu_e$ flux for $M_a=0.15\ M_\odot$ and $T_a=2.5$ MeV.
Dotted curve: black body distribution with the same luminosity
($1.1\times 10^{53}$~erg/s) and 
average energy ($13$~MeV), 
namely, with parameters $R_c=82$ km and $T_c=4.1$~MeV. 
\label{fig:pinch}}
\end{figure}

The energy spectrum of our model, for selected values of the astrophysical 
parameters, is given in Fig.~\ref{fig:pinch} (from\cite{mosca}). 
The spectrum is `pinched' and for the selected values of the astrophysical 
parameters it has a 
`pinching' factor of about $\sim 4$ 
(note that the  `pinching' is an output, not an input).

\subsection{Results}
\begin{figure}[h]
\caption{\footnotesize\it
LL model: antineutrino luminosity and average energy in the best fit
point.\label{fep1}}
\centerline{
\includegraphics[width=0.36\textwidth,height=0.3\textwidth]{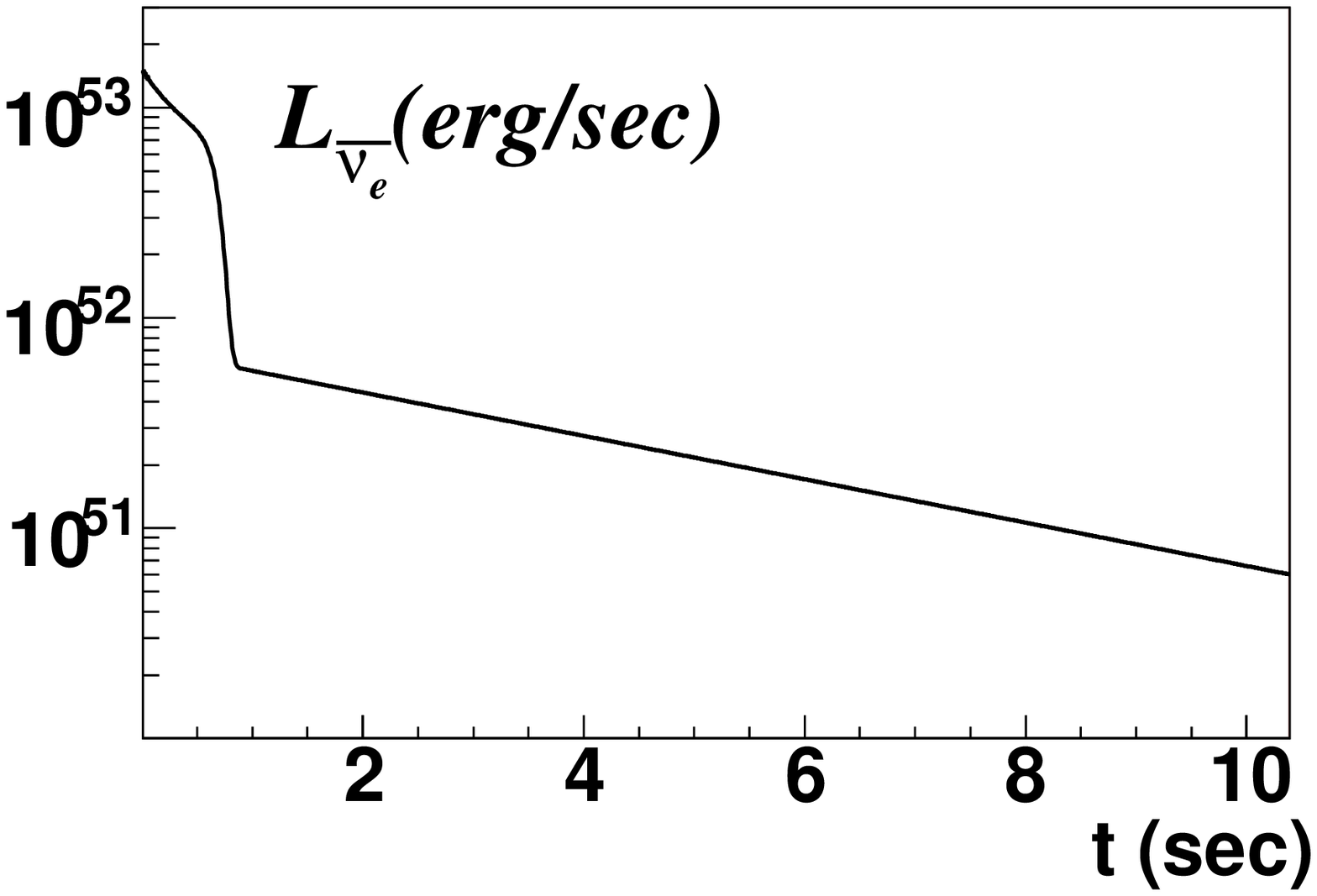}
\includegraphics[width=0.36\textwidth,height=0.3\textwidth]{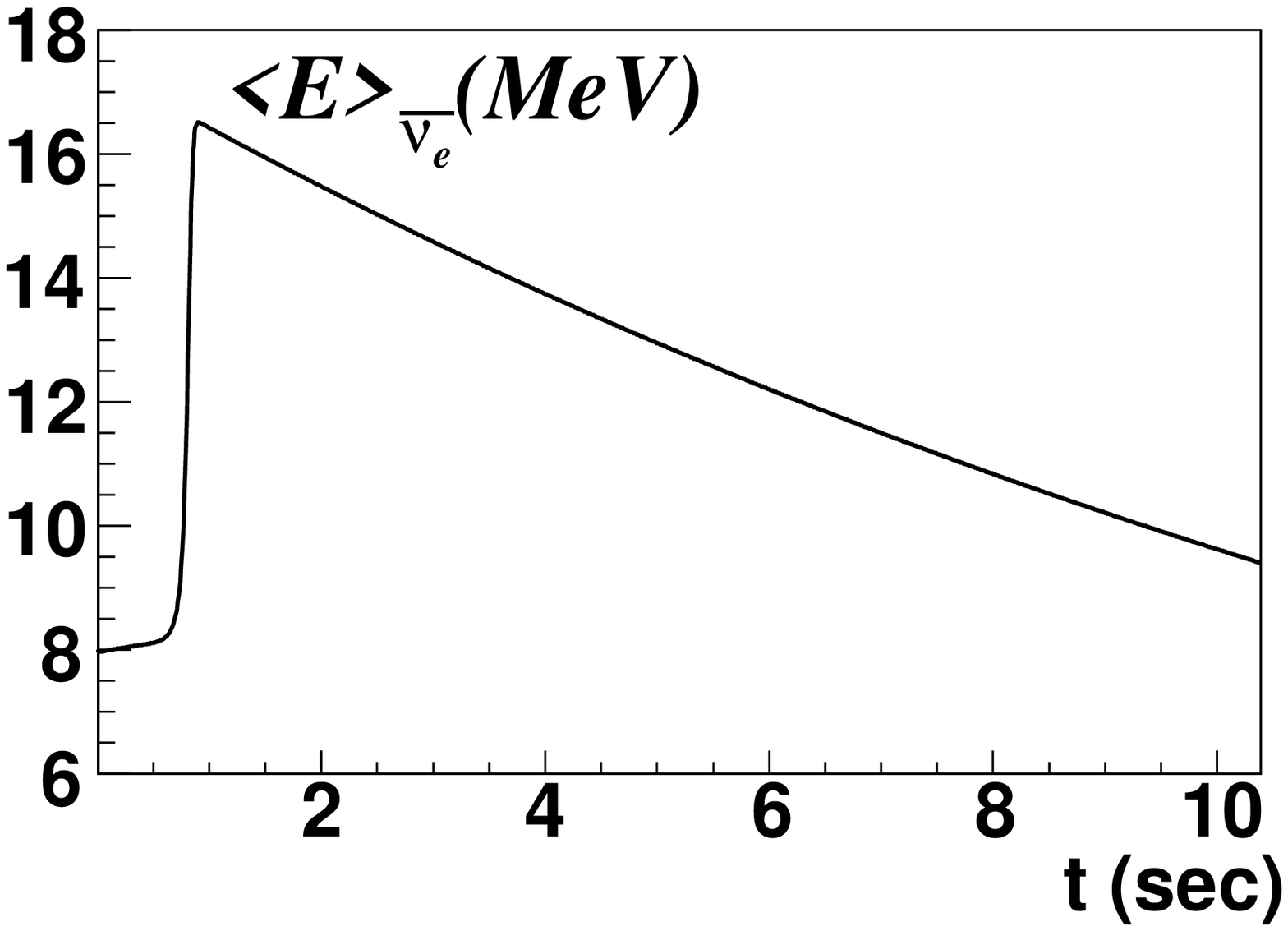}}
\end{figure}

Following the same procedure adopted for the one-component model, we 
obtain the values of the astrophysical parameters:
\paragraph{Lamb \& Loredo Model}
The best fit point is\cite{paglia}:
\begin{equation}
\begin{array}{ccc}
R_c=12\mbox{ km}, & T_c=5.5\mbox{ MeV}, & \tau_c=4.3\mbox{ s},\\ 
M_a={5.5\ M_\odot}, & T_a={1.5\mbox{ MeV}}, & 
\tau_a={0.7\mbox{ s}}.
\end{array}
\end{equation}
The big value of $M_a$ and the small value of $T_a$ are 
dictated by KII early events. The luminosities and average energies are 
given in Fig.~\ref{fep1}. 
The sharp transition at $t\sim \tau_a$ does not look very 
appealing, when compared with the result of a typical 
simulation.\cite{nosharp}

\begin{figure}[h]
\caption{\footnotesize\it
Pagliaroli et al.\  
model: antineutrino luminosity and average energy in the best fit point.
\label{fep2}
\label{fip}}
\centerline{
\includegraphics[width=0.36\textwidth,height=0.3\textwidth]{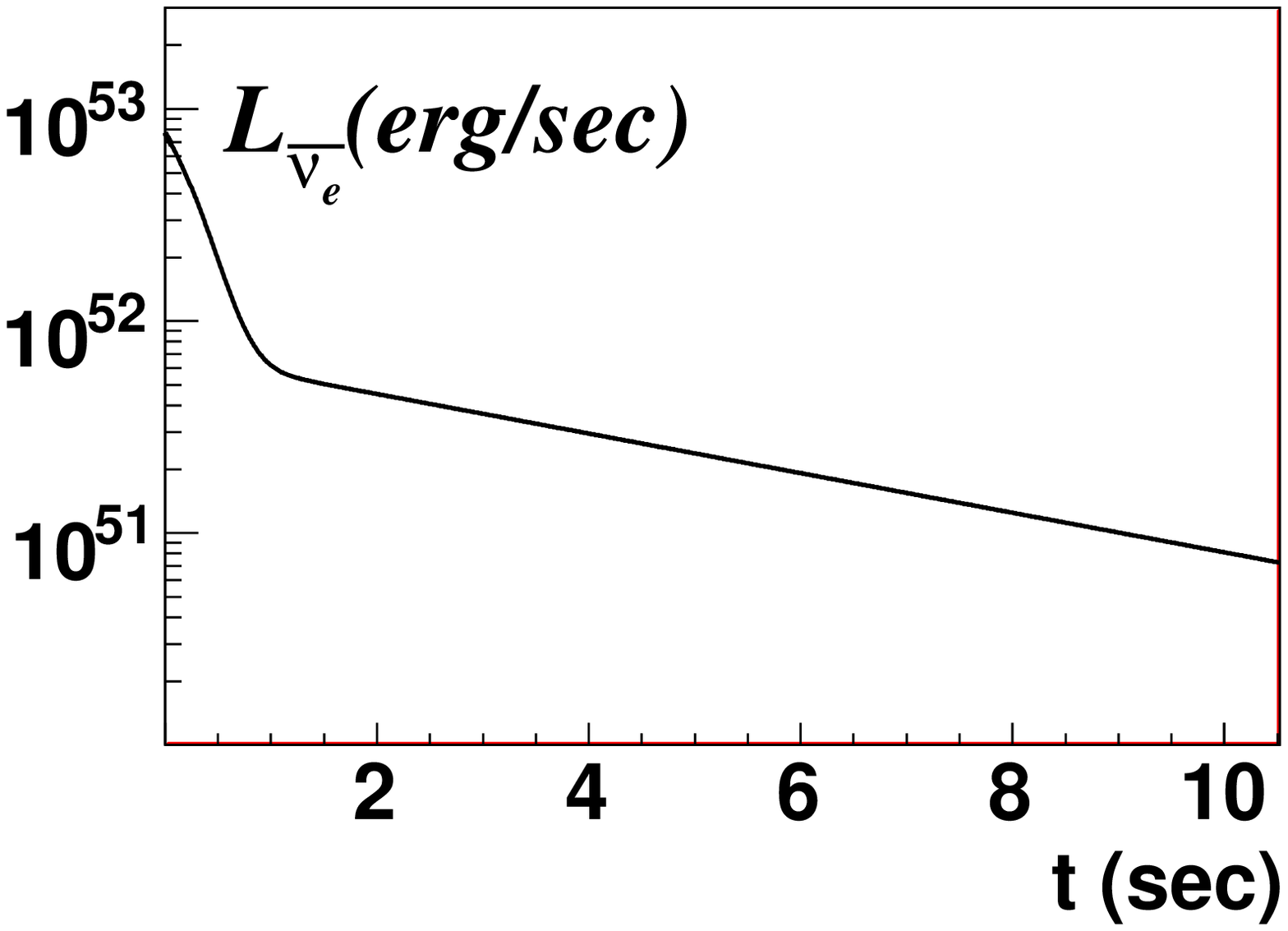}
\includegraphics[width=0.36\textwidth,height=0.3\textwidth]{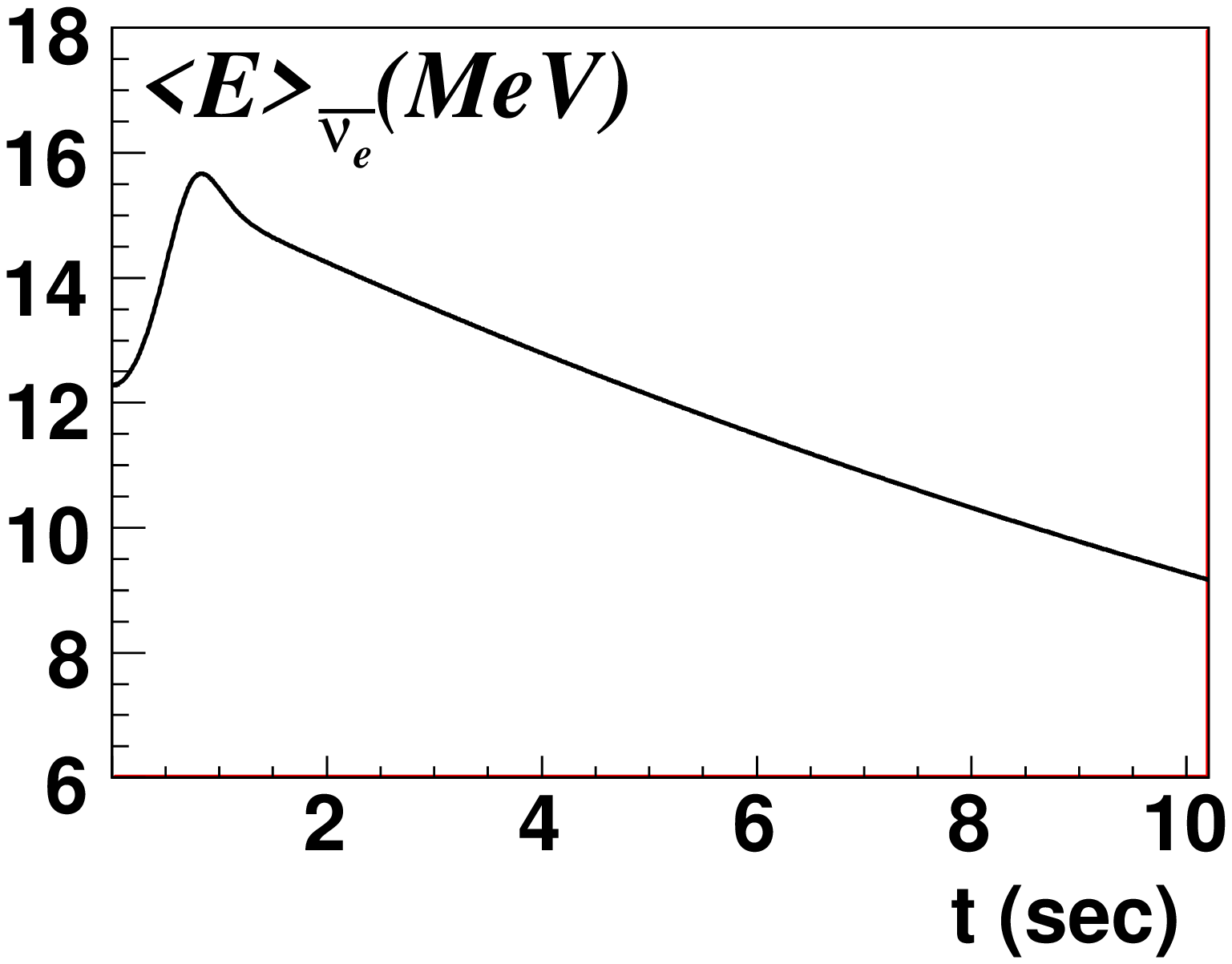}}
\end{figure}

\paragraph{Pagliaroli {\em et al.}\ Model}
The best fit point is\cite{paglia}:
\begin{equation}
\begin{array}{ccc}
R_c=16\mbox{ km}, & T_c=4.6\mbox{ MeV}, & \tau_c=4.7\mbox{ s},\\ 
M_a={0.2\ M_\odot}, & T_a={2.4\mbox{ MeV}}, & 
\tau_a={0.6\mbox{ s}}.
\end{array}
\end{equation}
Note that $M_a$ is much smaller and at the same time $T_a$
increased. 
The luminosity and average energy curves of Fig.~\ref{fep2} 
are much more regular and in much 
better agreement with the expectations.\cite{nosharp}

One may ask: what are the main reasons of the different result?
This is answered by Tab.~\ref{tab2} from\cite{paglia}, 
where we show the effect of the various
modification on the best fit point and we indicate the improvement of
the two-component models in comparison with the exponential cooling
model.
The two main differences of our model for antineutrino emission
with the model due to LL 
are 1)~that we assume a time dependent 
temperature $T_a$, that makes the average antineutrino energy continuous; 
2)~that, differently from LL, 
we postpone the occurrence of the cooling phase to 
the end of the accretion phase. In such a manner we have 
an approximately thermal spectrum 
at any time, as expected,\cite{thermos} 
and we remove the bi-modality that characterizes the 
first second of emission in the LL model.
It should be noted that these defects 
of the LL model have been noted already in 
Mirizzi and Raffelt\cite{raff}.

\begin{table}[h]
\begin{center}
\begin{tabular}{|c||c|c|c||c|c|c||c|}
\hline
\rm Subsequent    & $R_c$ & $T_c$ & $\tau_c$ & $M_a$ & $T_a$ & $\tau_a$ &
\rm {Signif.} \\
\rm improvement & \rm [km] & \rm [MeV] & \rm [s] & \rm [$M_\odot$] & \rm [MeV]
& [s]  &  {[\%]}\\
\hline\hline
\rm technical{\tiny\ ($\approx$LL)} & 12 & 5.5 & 4.3 & 5.6 & 1.5 & 0.7 & 
{99.8}\\
$T_a(t)$      & 14 & 5.0 & 4.8 & 0.8 & 1.8 & 0.7 & 
{98.9}\\
\rm time shift      & 14 & 4.9 & 4.7 & 0.1 & 2.4 & 0.6  &  
{98.0}\\
\rm oscillations    & 16 & 4.6 & 4.7 & 0.2 & 2.4 & 0.6  &  
{98.0}\\
\hline
\end{tabular}
\end{center}
\caption{\footnotesize\it Best-fit values of astrophysical parameters.
   Each line of this table is
  an incremental step toward the final improved parameterization. 
Last column shows the significance of the 2-component models in 
comparison (likelihood-ratio test, +3 d.o.f.) 
with the exponential cooling model. \label{tab2}}
\end{table}

The two-component models are better than
the exponential cooling model.
As a matter of fact, 
the LL-model would seem to fare better in describing 
the data, but it deviates more strongly 
from the expectations of the delayed scenario 
than our model.
Indeed, the LL model has less difficulty to 
account for the large difference between the energies detected by
Kamiokande-II and IMB in the first second, since it can  
ascribe these two datasets to two different 
but contemporaneous phases of emission. This bi-modality 
is forbidden by construction in our model.

\subsection{Errors in the Pagliaroli et al.\ Model}

The error on the astrophysical parameters can be
determined from the data\cite{paglia}:
\begin{equation}
\begin{array}{ll}
R_c=16^{+9}_{-5} \rm\ km,         &  M_a=0.22^{+0.68}_{-0.15}\ M_{\odot},\\
T_c=4.6^{+0.7}_{-0.6}\rm\ MeV,    &  T_a=2.4^{+0.6}_{-0.4} \rm\ MeV, \\
\tau_c=4.7^{+1.7}_{-1.2}\rm\ s,   &  \tau_a=0.55^{+0.58}_{-0.17}\rm\ s.
\end{array}
\end{equation}
We also show the values of the offset times: 
\begin{equation}
t^{\mbox{\tiny off}}_{{\mbox{\tiny KII}}}=0.^{+0.07}\rm\ s , \
t^{\mbox{\tiny off}}_{\mbox{\tiny IMB}}=0.^{+0.76}\rm\ s, \
t^{\mbox{\tiny off}}_{{\mbox{\tiny BAK}}}=0.^{+0.23}\rm\ s.
\end{equation}
We see that the 
limited statistics manifests itself in 
relatively large errors.

The marginal distributions 
for accretion parameters $M_a$ and $T_a$ of
$\bar{\nu}_e$ and for 
cooling parameters $R_c$ and $T_c$ of $\bar{\nu}_e$ 
with the complete emission model are given in Ref.\cite{paglia}.
Other solutions exist\cite{paglia} at $M_a\sim M_\odot$. 
Here we discarded them, considering that 
in the delayed scenario, $M_a$ should be a {\em fraction} of 
the outer core mass $M_{oc}\approx 0.6\ M_\odot$.

\begin{figure}[t]
\centerline{\includegraphics[width=0.55\textwidth,height=0.84\textwidth,angle=270]{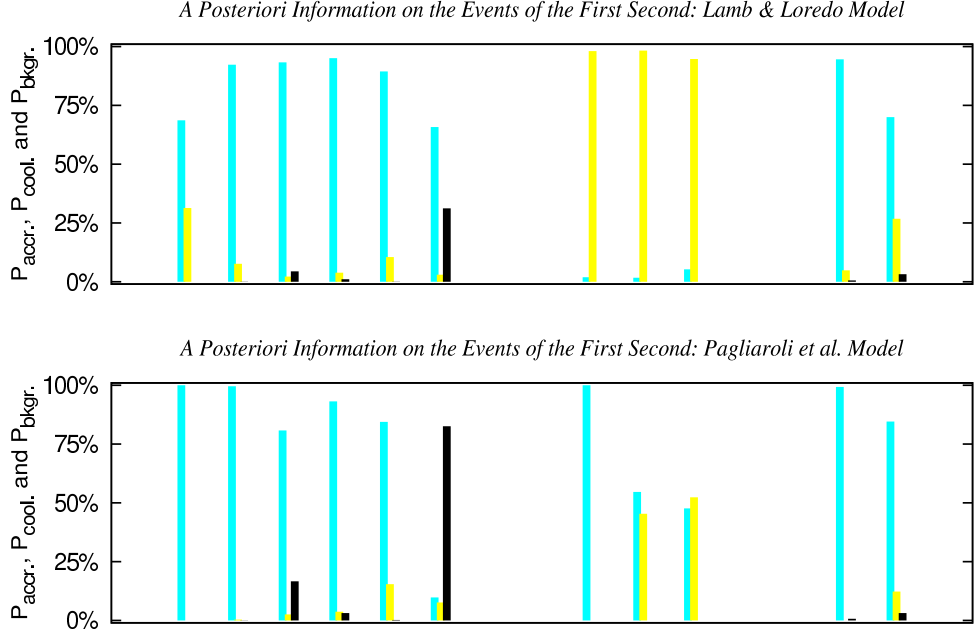}}
\caption{\footnotesize \it 
Events in the first second: 6 in KII, 3 in IMB, 2 in Baksan. 
First bar (light blue/gray), accretion probability;  
second bar (yellow/light gray), cooling probability;  
third bar (black), background probability.\label{appo}}
\end{figure}

\subsection{Meaning of the Individual Events}
Given a theoretical model for antineutrino emission, 
we can evaluate the meaning of each individual event. This is 
given in\cite{paglia} in table form and also in 
Fig.~\ref{appo} in graphical form
for the 11 events detected in the
first second and using the best fit models. 
We derive two principal conclusions:\\ 
$i)$ First, all 11 
events are {\em mostly}
due to signal and not to background, with the exception of the event 
number six of Kamiokande-II that is below the energy threshold of 7.5 MeV. 
The background probability of this event is larger in the model
by Pagliaroli {\em et al.}, due to the fact that it happened
0.686 s after the first one, namely, in the moment when the average 
energy of antineutrinos is maximal: see Fig.~\ref{fip}.\\
$ii)$ Second, we see from Fig.~\ref{appo} that in the LL model, the  
3 events of IMB  are attributed to cooling, 
the 6 events  of Kamiokande-II to accretion.
This is a manifestation of the bi-modality of the LL model
noted previously: the  
cooling and the accretion phases are assumed to be
contemporaneous and with very different average energies.
Thus the (low energy) events of Kamiokande-II are easily
explained  by accretion, 
the (high energy) ones of IMB by cooling. 
In the other model\cite{paglia} the spectrum is not bi-modal by construction.
Thus the 11 events are treated on equal footing 
and all of them (including IMB's) 
have a high probability to be due to accretion. 

Another 
manifestation of the bi-modality of the LL model is the 
great difference between the time integrated 
spectra of accretion and of cooling; 
see Fig.~2 of Ref.\cite{ifae}. This implies that the resulting 
total spectrum cannot described by a thermal distribution, even 
including a `pinching' factor.

\section{Conclusions}
Our study confirms earlier results with the 1-component model. 
The refined treatment of background, cross section, 
description of the scattering angle, 
inclusion of Baksan data, {\em etc.}\ do not
lead to important changes.

We confirm the results of Lamb \& {}Loredo in particular
the very strong evidence for accretion,  
when their 2-component model is adopted.

We discussed 
an improved 2-component model, where the 
average energy and luminosity curves are constrained to be continuous,
cooling follows accretion, oscillations 
(not very important {\em a posteriori})
are included.

The best fit of 
$\tau_a$, $M_a$, {\em etc.}\ are close to expectations;
the binding energy $2.2\times 10^{53}$ erg  
is lower than for the 1-component model; the 
evidence of accretion is not as strong as for the Lamb \& 
Loredo model, but still important.

\section{Acknowledgments}
F.V.~is very grateful to Milla Baldo-Ceolin 
for the very honorable invitation. 
We thank A.~Ianni and 
M.L.~Costantini for the stimulating 
collaboration, upon which this work is based.

\newpage
\appendix{: Derivation of the Likelihood Function}
We present here a derivation of the likelihood function 
that follows the one given in 
Appendix~A of LL, but notations and 
conclusions are somewhat different.
For a given model of the detector and of the antineutrino signal, 
we construct the likelihood:
\begin{equation}
{\cal L}=
\prod_{i=1}^N P_i \mbox{ where }
P_i=\left\{
\begin{array}{l}
P(d=0) \mbox{ if no events are detected in the $i$-th bin}\\
P(d(x_i)) \mbox{ if one event is detected in the $i$-th bin}
\end{array}
\right.
\end{equation}
the $i$=th time-bin being located in 
\begin{equation}
(i-1)\delta t\le t<i\delta t
\end{equation}
Of course, $d=0$ means that no event was detected in the given time bin.
$d(x)$ means that one event was detected in the time-bin
with coordinates (namely: energy, direction and position)
generically denoted as $x$.  
The product is on the number of time bins, $N$, each one 
of size $\delta t$; $T=N\delta t$. The number of 
bins is much larger than the number of detected events, 
$N\gg N_d$.
The two types of probability are calculated 
as follows:
\begin{equation}
\begin{array}{lcl}
P(d=0) & \simeq & 
P(d=0,e^+=0,b=0)+ \\
&&P(d=0,e^+=0,b=1)+\\
&&P(d=0,e^+=1,b=0) \\[1ex]
P(d(x)) & \simeq & 
P(d(x),e^+=0,b=1)+\\
&&P(d(x),e^+=1,b=0) 
\label{5}
\end{array}
\end{equation}
namely by considering the various cases, 
when the given time bin
contains 0 or 1 events due to background, $b$, 
or due to positrons produced in the detector, $e^+$.
This separation might appear cumbersome at first sight 
but it helps for comparing with Appendix~A of LL.
The notation $e^+=1$ ($b=1$) means that one positron 
(one background event) was in the detector in the given time bin, 
with some value of the the coordinates; similarly, 
$d=1$ means that one event was detected in the given time bin, 
with some value of the the coordinates.

Now we will expand the five terms in Eq.~\ref{5}
to order ${\cal O}(\delta t)$.

We begin with the simplest case when no event is present:
\begin{equation}
\begin{array}{l}
P(d=0,e^+=0,b=0) =\\   
P(d=0,e^+=0|b=0) \times P(b=0) =\\
P(d=0|e^+=0,b=0) \times P(e^+=0) \times P(b=0) =\\
1\times \exp(-S_+(t)\delta t) \times \exp(-B\delta t) =\\
1-\delta t (S_+(t)+B)
\label{b1}
\end{array}
\end{equation}
where we introduced the positron production rate $S_+(t)$, the (time 
independent) background rate $B$ and assumed Poisson statistics for 
both processes.

The next two cases have one background event but no positron
produced:
\begin{equation}
\begin{array}{l}
P(d=0,e^+=0,b=1) \equiv 0 \\[1ex]
P(d(x),e^+=0,b=1) \equiv \delta t B(x) 
\label{b2}
\end{array}
\end{equation}
This is basically a definition of what we mean by background: 
The background is a detected event that is not due to the 
signal.
Note that we are not attempting to describe the nature of the background
with this procedure, we are simply taking into account its existence.
However, having assumed 
that the average background is independent from the time of 
observation, there is practical way to obtain $B(x)$; 
this can be measured 
by the experimental collaborations 
in the instants of time when the signal is absent. 

The more complicate cases are the remaining two, 
when a positron is produced. Adopting the symbol $x_+$ for the coordinates
of the positron, we get:
\begin{equation}
\begin{array}{l}
P(d=0,e^+=1,b=0) = \\
\int dx_+ P(d=0,e^+(x_+),b=0) = \\
\int dx_+ P(d=0,e^+(x_+)|b=0)\times P(b=0) = \\
\int dx_+ P(d=0|e^+(x_+),b=0)\times P(e^+(x_+))\times P(b=0) = \\
\int dx_+ (1-\eta(x_+))\times S_+(t, x_+ ) \delta t 
\times 1 =\\
\delta t (S_+(t)- \int dx_+ \eta(x_+) S_+(t, x_+ )) = \\
\delta t ( S_+(t)- S(t) ) 
\label{b3}
\end{array}
\end{equation}
where we used 
$P(e^+(x_+))=S_+(t,x_+) \delta t \ \exp(-S_+(t,x_+) \delta t )$
and introduced the definition of detection efficiency,
that agrees with the one of LL:
\begin{equation}
P(d=0|e^+(x_+),b=0)=1-\eta(x_+)
\end{equation}

The last term to evaluate is:
\begin{equation}
\begin{array}{l}
P(d(x),e^+=1,b=0) = \\
\int dx_+ P(d(x),e^+(x_+),b=0) = \\
\int dx_+ P(d(x),e^+(x_+)|b=0)\times P(b=0) = \\
\int dx_+ P(d(x)|e^+(x_+),b=0)\times P(e^+(x_+))\times P(b=0) = \\
\int dx_+ \rho(x,x_+)\eta(x_+)\times S_+(t, x_+ ) \delta t \times 1 = \\
\delta t \int dx_+ \rho(x,x_+) \eta(x_+) S_+(t, x_+ )) = \\
\delta t S(t,x) 
\label{b4}
\end{array}
\end{equation}
where the last equality is simply the definition of $S(t,x)$.

We introduced the new function $\rho(x,y)$ according to: 
\begin{equation}
P(d(x)|e^+(x_+),b=0)\equiv \rho(x,x_+) \eta(x_+)
\end{equation}
The interpretation of this function is as follows. 
When we consider the probability that a positron 
is either detected or missed, we should get 1:
\begin{equation}
P(d=0|e^+(x_+),b=0)+ P(d=1|e^+(x_+),b=0)=1
\end{equation}
In formulae and with the definitions introduced above:
\begin{equation}
1-\eta(x_+)+\int dx \rho(x,x_+) \eta(x_+)=1
\end{equation}
that implies $\int dx \rho(x,x_+)=1$.
Thus, $\rho(x,x_+)$  can be thought of as the probability 
that the event produced with coordinates $x_+$ and 
eventually detected, will be observed 
with coordinates $x$:
$\rho(x,x_+)\equiv P(d(x)|e^+(x_+)\ ,\ e^+\mbox{ detected})$.
This function is occasionally called `smearing', since a 
positron with well-defined coordinates will be mapped 
by the detector response 
into  a wide set of coordinates of the detected event.
This concludes the calculation of the five terms in
Eq.~\ref{5}.\footnote{We remark here the main differences with LL:\\
The first one emerges when we compare 
the equation for $P(d(x)|e^+(x_+),b=0)$ with 
their Eqs.~(A.20) and (3.21) of LL. 
One concludes that the function ${\cal L}_i$ defined there corresponds 
to $\rho(x,x_+)$ but the factor $\eta(x_+)$ has been 
replaced with a term proportional to the  
step function $\Theta(\epsilon-\epsilon_0)$.\\
The second difference regards the meaning of background. 
For instance,  their eq.~(A.14) is given by the product of 
``background'' and ``efficiency''; 
this implies that some ``background'' events will be seen, 
some other will be missed. 
This conceptual scheme is appealing being completely 
analogous to the one adopted for the signal events, 
but in order to be practical, it would need a 
model for the ``background'' 
(namely, contamination of neutrons, of radon, 
electronic noise, etc)
along with a description of how the detector responds to 
these events (in particular, what are the 
``efficiencies'' and what are the ``smearing'' functions).  
We believe that it is much simpler to consider as 
background event (without quotation marks) 
an event that we detect in absence of signals
and this is the essence of the definition we described
previously.}

Plugging the results of 
Eqs.~\ref{b1}, \ref{b2}, \ref{b3} and \ref{b4} in Eq.~\ref{5} 
we find that the expression of the probabilities 
in the given time bin at the order $\delta t$:
\begin{equation}
\begin{array}{ll}
P(d=0)&=( 1-  \delta t (S_+(t)+B) ) + 0 + \delta t(S_+(t)-S(t))\simeq \\
& \simeq e^{-\delta t (B+S(t)) } \\[1ex]
P(d(x))& =\delta t ( B(x) +  S(t,x)) \simeq\\ 
& \simeq \delta t ( B(x) +  S(t,x)) e^{-\delta t (B+S(t)) }
\end{array}
\end{equation}
These can be introduced in the formula for the likelihood
finding:
\begin{equation}
{\cal L}=e^{-\int_0^T dt (B+S(t)) } \prod_{i=1}^{N_d} 
\delta t ( B(x_i) +  S(t_i,x_i))
\end{equation}
We can omit the multiplying, constant factors 
$\exp(-B T)$ and $(\delta t)^{N_d}$ without affecting the 
parameter estimations. We note finally that, owing to the definitions of 
$S(t)$ and $S(t,x)$, it is possible and occasionally convenient 
to define a detector-weighted interaction rate 
by including the efficiency $\eta(x_+)$ in the positron production rate:
\begin{equation}
R(t,x_+)\equiv S_+(t,x_+) \eta(x_+)
\end{equation}
The likelihood becomes
\begin{equation}
{\cal L}=\mbox{const.}\times
e^{-\int_0^T dt dx R(t,x) } \prod_{i=1}^{N_d} 
( B(x_i) +  \int dx \rho(x_i,x) R(t_i,x) )
\end{equation}

\appendix{: Details of the Pagliaroli {\em et al.}\ Model}

We estimate the theoretical parameters by the $\chi^2$:
\begin{equation}
\chi^2\equiv-2 \sum_{d=k,i,b} \log( {\cal L}_d ),
\end{equation}
where  ${\cal L}_d$ is the likelihood of any detector ($k,i,b$ are
shorthands for Kamiokande-II, IMB, Baksan). 
The likelihood of each of the 3 detectors is:
\begin{equation}
\begin{array}{l}
{\cal L}_d= e^{-f_d \int \!\!  R(t) dt} \times \prod^{N_d}_{i=1} e^{R(t_i) \tau_d  }\times \left[ B(E_i,\cos\theta_i) + \int dE \rho(E,E_i) 
R(t_i, E,\cos\theta_i)  \right]. 
\end{array}
\end{equation}
Setting $f_d=1$ and $\tau_d=0$ (that is adequate for 
Kamiokande-II and Baksan) and noting that the IBD reaction is only weakly 
directional (that  implies that the smearing on the angle can be 
in good approximation neglected) we see that this is equivalent 
to the likelihood discussed in detail just above.

The time dependent temperature is described as:
\begin{equation}
T_a(t)=T_i + (T_f-T_i) \left( \frac{t}{\tau_a} \right)^m
\!\! \mbox{ with }
\left\{
\begin{array}{l}
T_i=T_a \\
T_f=0.6\ T_c 
\end{array}
\right.
\end{equation}
with $m=1-2$. The drop of the number of neutrons with time is:
\begin{equation}
N_n(t)=\frac{Y_n}{m_n}\times M_a\times \left( \frac{T_a}{T_a(t)}
\right)^6 
\times \frac{j(t)}{1+t/0.5\mbox{ s}}
\end{equation}
where $Y_n=0.6$ and 
\begin{equation}j(t)=\exp[-(t/\tau_a)^2]\end{equation}
The time shift described as follows:
\begin{equation}
\Phi_{\bar\nu_e}(t) = 
\Phi_{a}(t) + (1-j(t))\times \Phi_{c}(t-\tau_a)
\end{equation}

Finally we give some details of the treatment of oscillations:
For \underline{normal} mass hierarchy (used in the calculation 
given in the main text) 
the survival probability and the
observed  $\bar\nu_e$ flux are:
\begin{equation}
\begin{array}{ll}
P=U_{e1}^2 ,\\[1ex]
\Phi_{\bar\nu_e}= P\ \Phi_{\bar\nu_e}^0 + (1-P)\ \Phi_{\bar\nu_\mu}^0,\\
\end{array} \label{normal}
\end{equation}
For \underline{inverted} mass hierarchy
$\nu-\nu$ interaction introduces a swap between the 
$\bar\nu_e$ and $\bar\nu_x$ so that
\begin{equation}
\begin{array}{ll}
P=U_{e1}^2 P_f + U_{e3}^2 (1-P_f),\\[1ex]
\Phi_{\bar\nu_e}= P\ \Phi_{\bar\nu_\mu}^0 + (1-P)\ \Phi_{\bar\nu_e}^0.\\
\end{array} \label{inverted}
\end{equation}
The earth matter effect is described by the PREM model.

\newpage

\appendix{: Discussion at the Workshop}

We report the discussion at 
the Workshop after the talk was delivered. 
This discussion emphasizes certain important points 
and it is (at least in our view) useful, 
but since we were not able to reproduce the exact words used
we apologize in advance (and take the full responsibility)
for possible misunderstanding.

\vskip5mm
{{\tt Q [P. Vogel]:} 
{\em Don't angular distributions disagree with expectations? 
In particular, IMB's? If you find the GOF is bad, 
you could doubt that such an analysis is legitimate.}} 

{\tt A:} {\small The GOF of the angular distribution 
is discussed in our previous publication,
PRD70:043006,2004. It is better than the conventional 5\%, 
so the problem is not so severe as one may feel. Our
re-evaluation of GOF and also the analysis 
uses the newly calculated cross section 
(PLB564:42-54,2003) and the published
angular bias of IMB\cite{imb}.\\ 
Again on  IMB, there is an important point to note: the
events of IMB cannot be elastic scattering events,
because they are not so directional. 
And even being ready to consider  
something exotic taking place, it
is hard to imagine a reaction that is forward peaked but too much, 
as needed to locate half of the events of IMB 
in the region where they are, 
$30^\circ <\theta<60^\circ$. All in all, 
the hypothesis that they are inverse beta decay events seems to be 
the most reasonable and conservative. \\
Other Authors explore the hypothesis that IMB analysis could
be biased, e.g., Malgin in Nuovo Cim.C21:317-329,1998 (that incidentally 
also proposes interesting remarks on the time distribution of the events): 
this is not the case of our work on SN1987A events.}

\vskip5mm

{{\tt Q [V. Berezinsky]:} 
{\em The role of rotation is important and should be included 
in the analysis.}}

{\tt A:} {\small We tried to avoid linking our analysis 
to a specific model, and we kept the parameters free. These 
results could be used as follows: Suppose that a certain model
with 
rotation produces an initial temperature of 
accretion $T_a$ much lower than the one of our best fit value; 
then, one could be entitled to conclude that this model with rotation 
is disfavored by the observations of SN1987A neutrino events.} 

\vskip5mm

{{\tt Q [V. Matveev]:} 
{\em Could some events be due to neutrino 
interactions with iron?}}

{\small 
{\tt A:} Yes the most complete analysis should take into account all the
interactions happening in the detectors. We tried to keep this
analysis as simple as possible and this is the reason why we mainly 
focused on the parameterization of the most important flux, namely,
the one of  
electron antineutrinos. As emphasized in the paper of Imshennik and
Ryazhskaya\cite{ir}, in order to assess the importance of the interactions with
iron, we would need to describe the electron neutrino flux, too. }

\end{document}